\documentclass[a4paper,twoside,twocolumn,aps,prl,superscriptaddress]{revtex4}
\usepackage{graphicx}
\usepackage{amsmath}

\begin{document}
\title{Climbing the Jaynes-Cummings Ladder\\ and Observing its $\sqrt{n}$ Nonlinearity in a Cavity QED System}
\author{J.~M.~Fink}
\affiliation{Department of Physics, ETH Zurich, CH-8093,
Zurich, Switzerland.}
\author{M.~G\"oppl}
\affiliation{Department of Physics, ETH Zurich, CH-8093,
Zurich, Switzerland.}
\author{M.~Baur}
\affiliation{Department of Physics, ETH Zurich, CH-8093,
Zurich, Switzerland.}
\author{R.~Bianchetti}
\affiliation{Department of Physics, ETH Zurich, CH-8093,
Zurich, Switzerland.}
\author{P.~J.~Leek}
\affiliation{Department of Physics, ETH Zurich, CH-8093,
Zurich, Switzerland.}
\author{A.~Blais}
\affiliation{D\'epartement de Physique, Universit\'e de
Sherbrooke, Sherbrooke, Qu\'ebec, J1K 2R1 Canada.}
\author{A.~Wallraff}
\affiliation{Department of Physics, ETH Zurich, CH-8093,
Zurich, Switzerland.}
\date{\today}

\begin{abstract}
\textbf{The already very active field of cavity quantum
electrodynamics (QED), traditionally studied in atomic systems
\cite{Raimond2001,Mabuchi2002,Walther2006}, has recently gained
additional momentum by the advent of experiments with semiconducting
\cite{Reithmaier2004,Yoshie2004a,Peter2005,Hennessy2007,Englund2007}
and superconducting
\cite{Wallraff2004b,Chiorescu2004a,Johansson2006} systems. In these
solid state implementations, novel quantum optics experiments are
enabled by the possibility to engineer many of the characteristic
parameters at will. In cavity QED, the observation of the vacuum
Rabi mode splitting is a hallmark experiment aimed at probing the
nature of matter-light interaction on the level of a single quantum.
However, this effect can, at least in principle, be explained
classically as the normal mode splitting of two coupled linear
oscillators \cite{Zhu1990}. It has been suggested that an
observation of the scaling of the resonant atom-photon coupling
strength in the Jaynes-Cummings energy ladder \cite{Walls1994} with
the square root of photon number $n$ is sufficient to prove that the
system is quantum mechanical in nature \cite{Carmichael1996}. Here
we report a direct spectroscopic observation of this characteristic
quantum nonlinearity. Measuring the photonic degree of freedom of
the coupled system, our measurements provide unambiguous, long
sought for spectroscopic evidence for the quantum nature of the
resonant atom-field interaction in cavity QED. We explore
atom-photon superposition states involving up to two photons, using
a spectroscopic pump and probe technique. The experiments have been
performed in a circuit QED setup \cite{Blais2004}, in which ultra
strong coupling is realized by the large dipole coupling strength
and the long coherence time of a superconducting qubit embedded in a
high quality on-chip microwave cavity. Circuit QED systems also
provide a natural quantum interface between flying qubits (photons)
and stationary qubits for applications in quantum information
processing and communication (QIPC) \cite{Nielsen2000}.}
\end{abstract}

\maketitle

\begin{figure}[b!]
\includegraphics[width=0.93 \columnwidth]{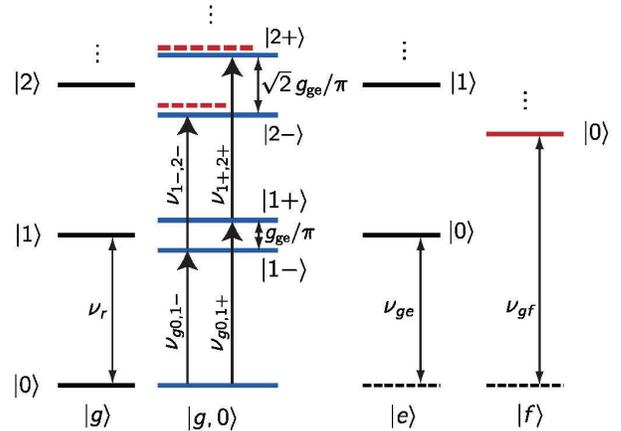}
\footnotesize
\caption{\textbf{Level diagram of a resonant ($\nu_r=\nu_{ge}$)
cavity QED system.} The uncoupled qubit states $|g\rangle,
|e\rangle$ and $|f\rangle$ from left to right and the photon states
$|0\rangle$, $|1\rangle$, ..., $|n\rangle$ from bottom to top are
shown. The dipole coupled dressed states are shown in blue and a
shift due to the $|f,0\rangle$ level is indicated in red. Pump
$\nu_{g0,1-}$, $\nu_{g0,1+}$ and probe $\nu_{1-,2-}$, $\nu_{1+,2+}$
transition frequencies are indicated accordingly.} \label{fig1}
\end{figure}

The dynamics of a two level system coupled to a single mode of an
electromagnetic field is described by the Jaynes-Cummings
Hamiltonian
\begin{equation}\label{jch}
\hat{\mathcal{H}}_{\textrm{0}}=\hbar\omega_{\textrm{ge}}\hat{\sigma}_{\textrm{ee}}
+\hbar\omega_\textrm{r} \hat{a}^\dagger \hat{a} +\hbar
g_{\textrm{ge}}(\hat{\sigma}^\dagger_{\textrm{ge}}\hat{a}
+\hat{a}^\dagger\hat{\sigma}_{\textrm{ge}}) \, .
\end{equation}
Here, $\omega_{\textrm{ge}}$ is the transition frequency between the
ground $|g\rangle$ and excited state $|e\rangle$ of the two level
system, $\omega_\textrm{r}$ is the frequency of the field and
$g_{\textrm{ge}}$ is the coupling strength between the two.
$\hat{a}^{\dagger}$ and $\hat{a}$ are the raising and lowering
operators acting on the photon number states $|n\rangle$ of the
field and $\hat{\sigma}_{\textrm{ij}}=|i\rangle\langle j|$ are the
corresponding operators acting on the qubit states. When the
coherent coupling rate $g_{\textrm{ge}}$ is larger than the rate
$\kappa$ at which photons are lost from the field and larger than
the rate $\gamma$ at which the two level system looses its
coherence, the strong coupling limit is realized. On resonance
($\omega_{\textrm{ge}} = \omega_{\textrm{r}}$) and in the presence
of $n$ excitations, the new eigenstates of the coupled system are
the symmetric $(|g,n\rangle+|e,n-1\rangle)/\sqrt{2}\equiv|n+\rangle$
and antisymmetric
$(|g,n\rangle-|e,n-1\rangle)/\sqrt{2}\equiv|n-\rangle$ qubit-photon
superposition states, see Fig.~\ref{fig1}. For $n=1$, these states
are equivalently observed spectroscopically as a vacuum Rabi mode
splitting
\cite{Thompson1992,Wallraff2004b,Boca2004,Reithmaier2004,Yoshie2004a,Peter2005,Hennessy2007,Englund2007}
or in time resolved measurements as vacuum Rabi oscillations
\cite{Brune1996,Varcoe2000,Bertet2002,Johansson2006} at frequency $2
g_{\rm{ge}}$. The Jaynes-Cummings model predicts a characteristic
nonlinear scaling of this frequency as $\sqrt{n} 2 g_{\rm{ge}}$ with
the number of excitations $n$ in the system, see Fig.~\ref{fig1}.
This quantum effect is in stark contrast to the normal mode
splitting of two classical coupled linear oscillators, which is
independent of the oscillator amplitude.

Since the first measurements of the vacuum Rabi mode splitting with,
on average, a single intra-cavity atom \cite{Thompson1992} it
remains a major goal to clearly observe this characteristic
$\sqrt{n}$ nonlinearity spectroscopically to prove the quantum
nature of the interaction between the two-level system and the
radiation field \cite{Zhu1990,Carmichael1996,Thompson1998}. In time
domain measurements of vacuum Rabi oscillations, evidence for this
$\sqrt{n}$ scaling has been found with circular Rydberg atoms
\cite{Brune1996} and superconducting flux qubits
\cite{Johansson2006} interacting with weak coherent fields. Related
experiments have been performed with one and two-photon Fock states
\cite{Varcoe2000,Bertet2002}.
We now observe this nonlinearity directly using a scheme similar to
the one suggested in Ref.~\cite{Thompson1998} by pumping the system
selectively into the first doublet $|1\pm\rangle$ and probing
transitions to the second doublet $|2\pm\rangle$. This technique
realizes efficient excitation into higher doublets at small intra
cavity photon numbers avoiding unwanted a.c.~Stark shifts occurring
in high drive and elevated temperature experiments.

In a different regime, when the qubit is detuned by an amount
$|\Delta|=|\omega_\textrm{ge}-\omega_\textrm{r}|\gg g_{ge}$ from the
cavity, photon number states and their distribution have recently
been observed using dispersive quantum non-demolition measurements
in both circuit QED \cite{Schuster2007a} and Rydberg atom
experiments \cite{Guerlin2007}.

\begin{figure}[b!]
\includegraphics[width=1.00 \columnwidth]{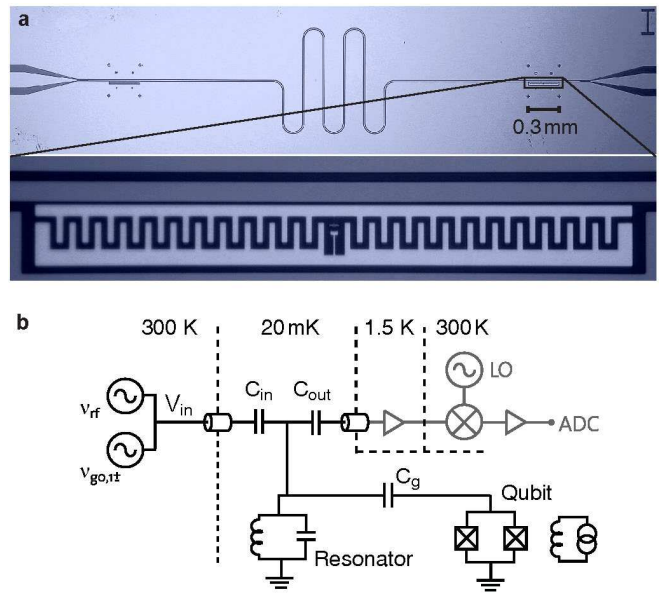}
\footnotesize
\caption{\textbf{Sample and experimental setup.} {\bf a},~Optical
images of the superconducting coplanar waveguide resonator (top)
with the transmon type superconducting qubit embedded at the
position indicated. On the bottom, the qubit with dimensions
$300\times30~\mu\textrm{m}^2$ close to the center conductor is
shown. {\bf b},~Simplified circuit diagram of the experimental
setup, similar to the one used in Ref.~\cite{Wallraff2004b}. The
qubit is capacitively coupled to the resonator through $C_{\rm{g}}$
and the resonator, represented by a parallel LC circuit, is coupled
to input and output transmission lines via the capacitors
$C_{\rm{in}}$ and $C_{\rm{out}}$.  Using ultra low noise amplifiers
and a down-conversion mixer, the transmitted microwave signal is
detected and digitized.} \label{fig2}
\end{figure}

In our experiments, in the resonant regime a superconducting qubit
playing the role of an artificial atom is strongly coupled to
photons contained in a coplanar waveguide resonator in an
architecture known as circuit QED \cite{Blais2004,Wallraff2004b}. We
use a transmon \cite{Koch2007,Schreier2007}, which is a
charge-insensitive superconducting qubit design derived from the
Cooper pair box (CPB) \cite{Bouchiat1998}, as the artificial atom.
Its transition frequency is given by
$\omega_{\textrm{ge}}/2\pi\simeq\sqrt{8
E_\textrm{C}E_\textrm{J}(\Phi)}$ with the single electron charging
energy $E_\textrm{C}\approx0.4 \, ~\textrm{GHz}$, the flux
controlled Josephson energy
$E_\textrm{J}(\Phi)=E_{\textrm{J,max}}|\cos{(\pi
\Phi/\Phi_\textrm{0})}|$ and $E_{\textrm{J,max}}\approx53.5 \,
\textrm{GHz}$, as determined in spectroscopic measurements. The
cavity is realized as a coplanar resonator with bare resonance
frequency $\nu_\textrm{r}\approx6.94 \, \textrm{GHz}$ and decay rate
$\kappa/2\pi\approx0.9 \, \textrm{MHz}$. Optical images of the
sample are shown in Fig.~\ref{fig2}a. The large dimension of the
qubit in the quasi one dimensional resonator layout provides a very
large dipole coupling strength $g_{\textrm{ge}}$. A simplified
electrical circuit diagram of the setup is shown in
Fig.~\ref{fig2}b.

\begin{figure}[b!]
\includegraphics[width=0.95 \columnwidth]{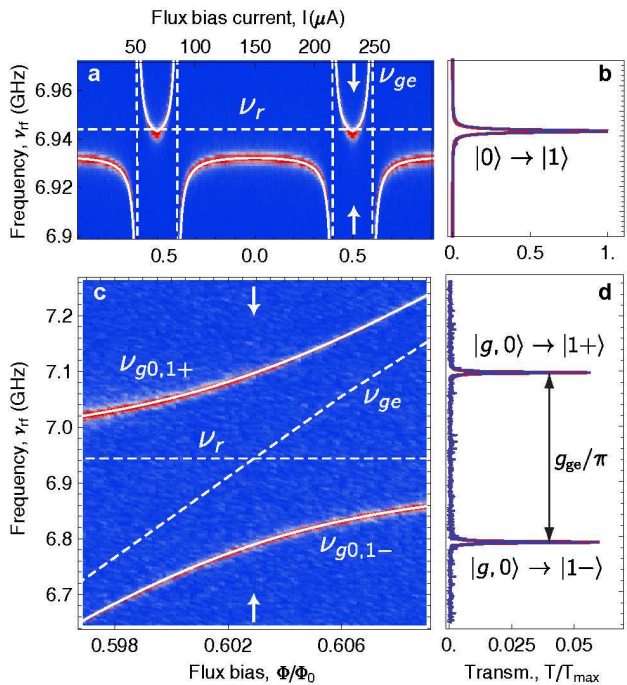}
\footnotesize
\caption{\textbf{Vacuum Rabi mode splitting with a single
photon.} {\bf a},~Measured resonator transmission spectra versus
external flux $\Phi$. Blue indicates low and red high transmission
$T$. The solid white line shows dressed state energies as obtained
numerically and the dashed lines indicate the bare resonator
frequency $\nu_\textrm{r}$ as well as the qubit transition frequency
$\nu_{\textrm{ge}}$. {\bf b},~Resonator transmission $T$ at
$\Phi/\Phi_{0}=1/2$ as indicated with arrows in panel \textbf{a}, with a
Lorentzian line fit in red. {\bf c},~Resonator transmission $T$
versus $\Phi$ close to degeneracy. {\bf d},~Vacuum Rabi mode
splitting at degeneracy with Lorentzian line fit in red.}
\label{fig3}
\end{figure}

The system is prepared in its ground state $|g,0\rangle$ by cooling
it to temperatures below $20\, \textrm{mK}$ in a dilution
refrigerator. We then probe the energies of the lowest doublet
$|1\pm\rangle$ measuring the cavity transmission spectrum $T$ and
varying the detuning between the qubit transition frequency
$\nu_{ge}$ and the cavity frequency $\nu_r$ by applying a magnetic
flux $\Phi$, see Fig.~\ref{fig3}a.  The measurement is performed
with a weak probe of power $P \approx -137~\textrm{dBm}$ applied to
the input port of the resonator
populating it with a mean photon number of $\bar{n}\approx 1.6$ on
resonance when the qubit is maximally detuned from the resonator.
$P$ is calibrated in a dispersive a.c.~Stark shift measurement
\cite{Schuster2005}. At half integers of a flux quantum
$\Phi_\textrm{0}$, the qubit energy level separation
$\nu_{\textrm{ge}}$ approaches zero. At this point the bare
resonator spectrum  peaked at the frequency $\nu_r$ is observed, see
Fig.~\ref{fig3}b. We use the measured maximum transmission amplitude
to normalize the amplitudes in all subsequent measurements. At all
other detunings $|\Delta| \gg g_{ge}$ the qubit dispersively shifts
\cite{Koch2007} the cavity frequency $\nu_\textrm{r}$ by $\chi
\simeq -g_{ge}^2 E_\textrm{C}/(\Delta (\Delta-E_\textrm{C}))$.

\begin{figure}[b!]
\includegraphics[width=0.85 \columnwidth]{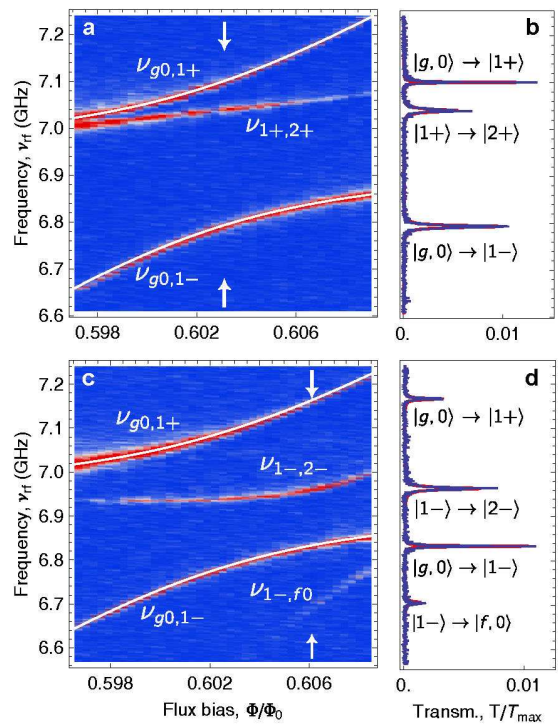}
\footnotesize
\caption{\textbf{Vacuum Rabi mode splitting with two photons.} {\bf
a},~Cavity transmission $T$ as in Fig.~\ref{fig3} with an additional
pump tone applied to the resonator input at frequency $\nu_{g0,1+}$
populating the $|1+\rangle$ state. {\bf b}, Spectrum at $\Delta=0$,
indicated by arrows in \textbf{a}. {\bf c},~Transmission $T$ with a pump
tone applied at $\nu_{g0,1-}$ populating the $|1-\rangle$ state.
{\bf d}, Spectrum at $\Phi/\Phi_0\approx0.606$ as indicated by
arrows in \textbf{c}.} \label{fig4}
\end{figure}

Measuring cavity transmission $T$ as a function of flux bias $\Phi$
in the anti-crossing region
yields transmission maxima at frequencies corresponding to
transitions to the first doublet $|1\pm\rangle$ in the
Jaynes-Cummings ladder as shown in Fig.~\ref{fig3}c. On resonance
($\Delta = 0$), we extract a coupling strength of
$g_{\textrm{ge}}/2\pi = 154 \, \rm{MHz}$, see Fig.~\ref{fig3}d,
where the linewidth of the individual vacuum Rabi split lines is
given by $\delta_{\nu 0}\approx 2.6 \, \rm{MHz}$. This
corresponds to a transmission peak separation $g_{\textrm{ge}}/\pi$
of over $100$ linewidths $\delta_{\nu 0}$, clearly demonstrating
that the strong coupling limit is realized \cite{Wallraff2004b,Schoelkopf2008}.
Solid white lines in
Figs.~\ref{fig3} (and \ref{fig4}) are numerically calculated dressed
state frequencies with the qubit and resonator parameters as stated
above, being in excellent agreement with the data. For the
calculation, the qubit Hamiltonian is solved exactly in the charge
basis. The qubit states $|g\rangle$ and $|e\rangle$ and the flux
dependent coupling constant $g_{\textrm{ge}}$ are then incorporated
in the Jaynes-Cummings Hamiltonian Eq.~(\ref{jch}). Its numeric
diagonalization yields the dressed states of the coupled system
without any fit parameters.

In our pump and probe scheme we first determine the exact energies
of the first doublet $|1\pm\rangle$ at a given flux $\Phi$
spectroscopically. We then apply a pump tone
at the fixed frequency $\nu_{g0,1-}$ or $\nu_{g0,1+}$ to populate
the respective first doublet state $|1\pm\rangle$. A probe tone of
the same power is then scanned over the frequency range of the
splitting. This procedure is repeated for different flux controlled
detunings. The transmission at the pump and probe frequencies is
spectrally resolved in a heterodyne detection scheme.

Populating the symmetric state $|1+\rangle$, we observe an
additional transmission peak at a probe tone frequency that varies
with flux, as shown in Fig.~\ref{fig4}a. This peak corresponds to
the transition between the symmetric doublet states $|1+\rangle$ and
$|2+\rangle$ at frequency $\nu_{1+,2+}$. Similarly, in
Fig.~\ref{fig4}c where the antisymmetric state $|1-\rangle$ is
populated we measure a transmission peak that corresponds to the
transition between the two antisymmetric doublet states $|1-\rangle$
and $|2-\rangle$ at frequency $\nu_{1-,2-}$.
The transmission spectra displayed in Figs.~\ref{fig4}b and d
recorded at the values of flux indicated by arrows in
Figs.~\ref{fig4}a and c show that the distinct transitions between
the different doublets are very well resolved with separations of
tens of linewidths.
Transitions between
symmetric and antisymmetric doublet states are not observed in this experiment, because the flux-dependent transition matrix elements squared are on average smaller by a factor of $10$ and $100$ for transitions $|1+\rangle \rightarrow |2-\rangle$ and $|1-\rangle \rightarrow |2+\rangle$, respectively, than the corresponding matrix elements between states of the same symmetry.

\begin{figure}[t!]
\includegraphics[width=0.7 \columnwidth]{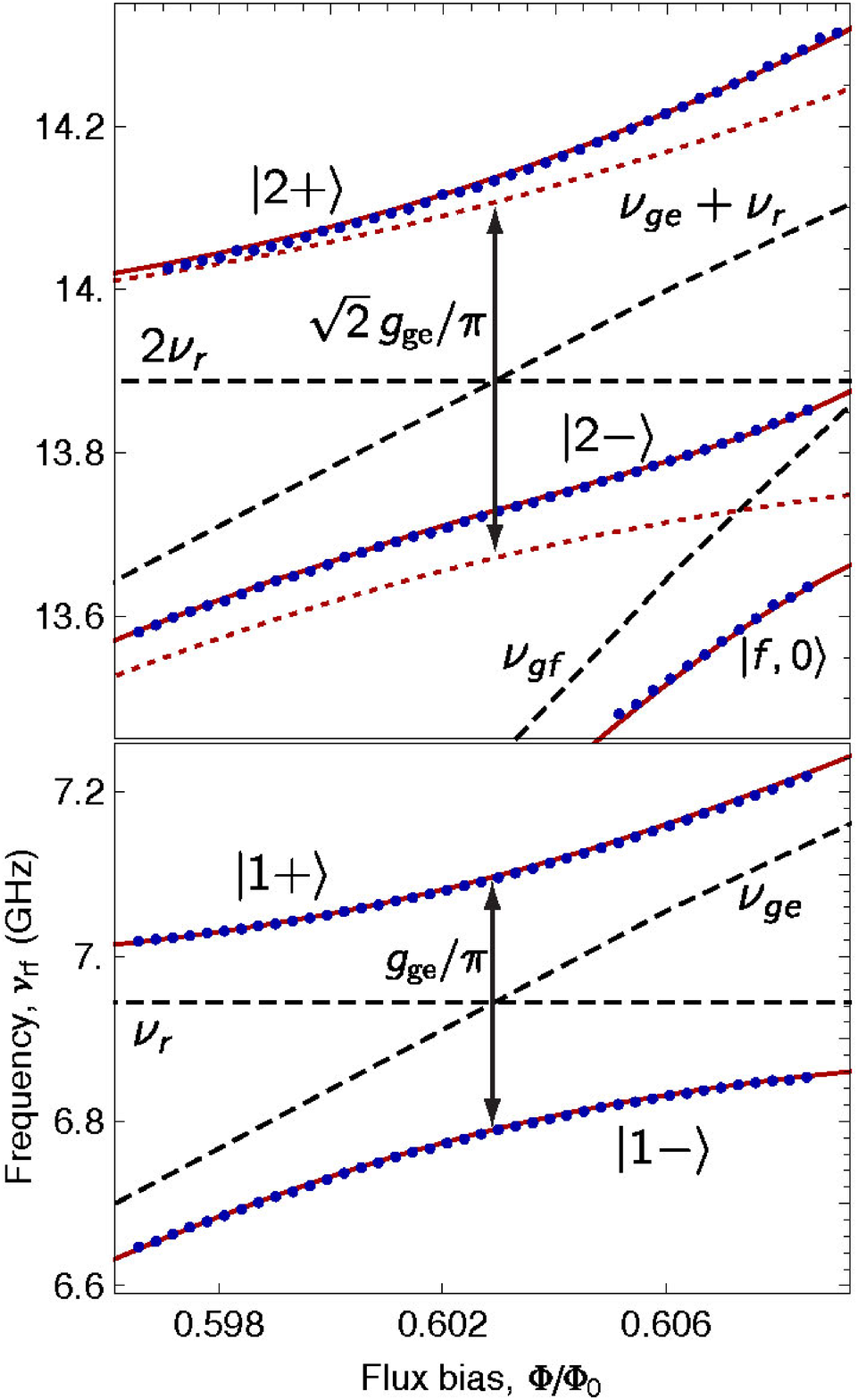}
\footnotesize
\caption{\textbf{Experimental dressed state energy levels.} Measured
dressed state energies (blue dots) reconstructed by summing pump and
probe frequencies, compared to the calculated uncoupled cavity and
qubit levels (dashed lines), the calculated dressed state energies
in the qubit two-level approximation (dotted) and to the
corresponding calculation including the third qubit level (solid red
lines).} \label{fig5}
\end{figure}

The energies of the first doublet $|1\pm\rangle$, split by
$g_{\textrm{ge}}/\pi$ on resonance, are in excellent agreement with
the dressed states theory (solid red lines) over the full range of
flux $\Phi$ controlled detunings, see Fig.~\ref{fig5}. The absolute
energies of the second doublet states $|2\pm\rangle$ are obtained by
adding the extracted probe tone frequencies $\nu_{1-,2-}$ and
$\nu_{1+,2+}$ to the applied pump frequencies $\nu_{g0,1-}$ or
$\nu_{g0,1+}$, see blue dots in Fig.~\ref{fig5}. For the second
doublet, we observe two peaks split by $1.34~g_{\textrm{ge}}/\pi$ on
resonance, a value very close to the expected $\sqrt{2}\sim1.41$.
This small frequency shift can easily be understood, without any fit
parameters, by taking into account a third qubit level
$|f,0\rangle$ which is at frequency $\nu_{\textrm{gf}}\simeq 2
\nu_{\textrm{ge}}-E_\textrm{C}$ for the transmon type qubit
\cite{Koch2007}, just below the second doublet states
$|2\pm\rangle$. In order to find the energies of the dressed states
in the presence of this additional level we diagonalize the
Hamiltonian
$\hat{\mathcal{H}}=\hat{\mathcal{H}}_{\textrm{0}}+\hat{\mathcal{H}}_{\textrm{1}}$,
where
$\hat{\mathcal{H}}_{\textrm{1}}=\hbar\omega_{\textrm{gf}}\hat{\sigma}_{\textrm{ff}}
+\hbar g_{\textrm{ef}}(\hat{\sigma}^\dagger_{\textrm{ef}}\hat{a}
+\hat{a}^\dagger\hat{\sigma}_{\textrm{ef}}) $ and
$g_{\textrm{ef}}/2\pi\approx 210 \, \rm{MHz}$ (obtained from exact
diagonalization)
denotes the coupling of the $|e\rangle$ to $|f\rangle$ transition to
the cavity. The presence of the $|f,0\rangle$ level is observed to
shift the antisymmetric state $|2-\rangle$, being closer in
frequency to the $|f,0\rangle$ state, more than the symmetric state
$|2+\rangle$, see Figs.~\ref{fig1} and \ref{fig5}, leading to the
small difference of the observed splitting from $\sqrt{2}$. The
$|f,0\rangle$ state, being dressed by the states $|g,2\rangle$ and
$|e,1\rangle$, is also directly observed in the spectrum via the
transition $|1-\rangle \rightarrow |f,0\rangle$ at frequency
$\nu_{1-,f0}$, see Fig.~\ref{fig4}c. This is in excellent agreement
with the dressed states model, see Fig.~\ref{fig5}. For comparison
the dressed states split by $\sqrt{2}g_{\textrm{ge}}/\pi$ in the
absence of the $|f,0\rangle$ state are shown as dotted red lines in
Fig.~\ref{fig5}.

Our experiments clearly demonstrate the quantum non-linearity of a
system of one or two photons strongly coupled to a single artificial
atom in a cavity QED setting. Both symmetric and antisymmetric
superposition states involving up to two photons are resolved by
many tens of linewidths. Recently, signatures of the $|2-\rangle$ state have
also been observed spectroscopically in an independent work on
optical cavity QED \cite{Schuster2008}. We have also observed that
higher excited states of the artificial atom can induce energy
shifts in the coupled atom-photon states. These shifts should also
be observable in time resolved measurements of Rabi-oscillations
with photon number states. In this circuit QED system, excited
states $|n\pm\rangle$ with $n > 2$ are also observable both by
pumping the system with thermal photons and by applying strong
coherent drive fields inducing multi-photon transitions. The
observed very strong nonlinearity on the level of single or few quanta could be used for the realization of a single photon
transistor, parametric down-conversion, and for the generation and
detection of individual microwave photons.

\begin{acknowledgments}
We thank L.~S.~Bishop, J.~M.~Chow, T.~Esslinger, L.~Frunzio,
A.~Imamo\u{g}lu, B.~R.~Johnson, Jens Koch, R.~J.~Schoelkopf and
D.~I.~Schuster for valuable discussions. This work was supported by
SNF and ETHZ. P.~J.~L.~was supported by the EU with a MC-EIF.
A.~B.~was supported by NSERC, CIFAR and FQRNT.
\end{acknowledgments}

\end{document}